\documentclass[prl,twocolumn,letterpaper,amsfonts]{revtex4}
\usepackage{color}
\usepackage{graphicx}

\usepackage{amsthm}
\usepackage{amsmath}
\usepackage{amsfonts}

\usepackage{amssymb}

\usepackage{bm}
 \usepackage{lineno}
 
 \usepackage{hyperref}

\begin{document}

\bibliographystyle{apsrev}

\newcommand{\bra}[1]{\ensuremath{\left \langle #1 \right |}}
\newcommand{\ket}[1]{\ensuremath{\left | #1 \right \rangle}}
\newcommand{\braket}[2]{\ensuremath{\left \langle #1 \right | \left. #2 \right \rangle}}
\newcommand{\tr}{\ensuremath{\mbox{Tr}}}
\newcommand{\R}{{\bf r}H}
\newcommand{\V}{\widehat{{\bf r}H}}
\newcommand{\U}{{\bf t}H}
\renewcommand{\H}{\ensuremath{\mathcal{H}}}
\newcommand{\var}{\ensuremath{ \mbox{var}}}
\newcommand{\hvar}[1]{\ensuremath{ \hat{\mbox{var}} \left ( #1 \right )}}
\newcommand\Prefix[3]{\vphantom{#3}#1#2#3}

\title{Democratizing University Research}
\author{Nick S. Jones$^1$, Oscar Ces$^2$}

\affiliation{$^1$Department of Mathematics, Imperial College London, London SW7 2AZ, United Kingdom}
\affiliation{$^2$Department of Chemistry, Imperial College London, Molecular Sciences Research Hub, London W12 0BZ, United Kingdom}


\begin{abstract}
We detail an experimental programme we have been testing in our university. Our Advanced Hackspace attempts to give all members of the university, from students to technicians, free access to the means to develop their own interdisciplinary research ideas, with resources including access to specialized fellows and biological and chemical hacklabs. We assess the aspects of our programme that led to our community being one of the largest collectives in our university and critically examine the successes and failures of our trial programmes. We supply metrics for assessing progress and outline challenges. We conclude with future directions that advance interdisciplinary research empowerment for all university members.
\end{abstract}

\maketitle

There is a huge opportunity to equalise access to the means of research in universities: the bulk of university research is performed by research groups led by faculty members, yet all members of universities (from undergraduates to technicians) can contribute to human enquiry and progress. Given typical academic-to-non-academic ratios \cite{russell} this equalization could lead to an order of magnitude change in the number of researchers inside universities. Arguably the roots of the word university, speaking to a body or corporation, acknowledge a coherent ability for the members of the university to participate in its function. A coarse view of many modern universities, however, is that academics teach and research and the majority of the rest of the university body is taught; but the distinction between research and teaching is delicate since research itself is a perpetual learning process: teaching can be achieved by performing research and research-led undergraduate courses can yield improved outcomes \cite{Nap}. Most academics are passionate about research and learning and want to see others enjoy and exploit this process; but despite the individual sentiments of academics, universities make it challenging for the majority of members to autonomously pursue their own research ideas. The challenges in democratizing research practice are closely linked to the challenges of interdisciplinary research. These challenges occur at levels of both physical and psychological accessibility: members of a biochemistry department might (sometimes) have access to their own departmental resources but will often have no access to electronics or machine shops; further, at the level of mind-set, scientists, young and old, can lack confidence in unfamiliar technologies from other disciplines even if the core skills can be rapidly acquired. Our challenge is then: how can we create courageous, practically empowered researchers, at a whole-university scale?

We believe these concerns are timely. The compelling, convergence \cite{conv} case for deeper integration of disciplines and technologies for the goal of exploring biomedicine requires an open university culture: supporting advances from cellular-bionics \cite{Bionics} to precision healthcare \cite{Colijn}. This requires a complementary perspective to the maker/fablab/university-makerspace movement \cite{Mikhak, Barrett, Forest} going beyond offering the baseline means of fabrication and design to offering the interdisciplinary means of modern scientific experimentation from genetic engineering to molecular synthesis. We stress that an experiment imaging a single cell `makes' nothing but is a vital scientific activity. Sequencing, 3d printing and maker-spaces should be interesting to universities because of what they signal: cheap and/or deskilled technologies can lower the barriers to access, making it increasingly feasible to perform cheap research on a large scale. While there is excitement about the possibilities of science-based student-start-ups we suggest that exploration, learning and experimentation are primitives of the university experience which reach beyond the preserve of entrepreneurs; however, investors and industrial support could be critical for substantial increases in the amount of research Universities deliver. As noted, there is accumulating evidence that active learning can be beneficial and that research projects can enhance student learning \cite{Freeman, Deslauriers, Nap}; we suggest, like others, that while exploration and experimentation are vital learning tools for changing outcomes in tests, it is the confidence and competence in autonomous enquiry and autonomous learning that research experiences provide \cite{Adedokun} that are a key transferable skill. We further suggest that, for all members, a period in university and afterwards could be an opportunity to directly contribute to human knowledge and progress: \emph{autonomous research by all university members can be a  goal in itself rather than exclusively a means to an educational end}. We thus consider a system that allows all university members autonomous access to the means to perform research outside either faculty-led research projects or faculty-led courses.

We have observed two barriers to empowering the autonomous research of non-academic members of universities: access to laboratory resources \cite{Speck} and the confidence required to experiment and self-teach. With the goal of providing a case-study for others we outline our interdiscipliary programme to address this, the Advanced Hackspace, and critically assess which parts of our trial programmes succeeded and failed.

\vspace{-3mm}

\subsection{A simple framework for democratizing university research:}

\vspace{-3mm}

\begin{figure}[h!]
\includegraphics[angle = 0, width =8.5cm, keepaspectratio=true]{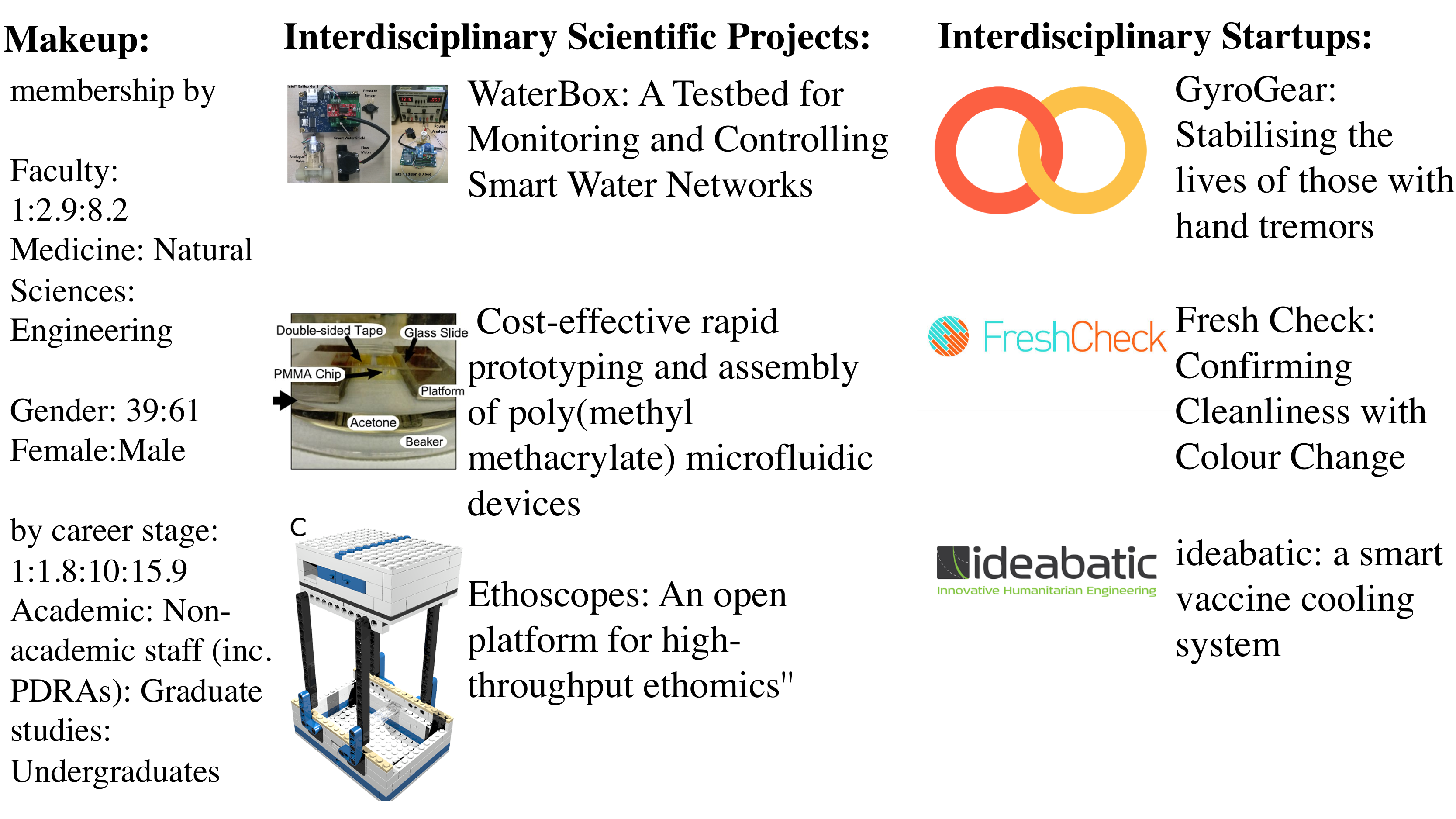}
\caption{\label{figure1}
Accelerating interdisciplinary research by empowering all university members: showing (left) cross-departmental and career stage mixing; (centre) example scientific publications \cite{waterbox,chemo,etho}; (right) interdisciplinary start-ups.}
\end{figure}

The Advanced Hackspace was started in May 2014 and, before all university members became auto-enrolled (for free) in Oct. 2017, we had 2,300 members \cite{ICAH}. We are staffed by a manager, administrator, community officer, commercial engagement officer and $\sim$6 Hackspace Fellows/Associates with backgrounds in either academic or industrial R\&D.  
Our mission is to be the best place in the world to turn ideas into a working reality: aiming to make it as easy for any member of the university to develop an idea as it is for an academic. Our team's principle operational goal is to safely and sustainably support the idea development of our members and their idea development community. We have three spaces covering $>$8000 sq ft: 2 shared with different academic departments (the first with Electrical Engineering and the second with Chemistry is being assembled) and 1 new hub-space. Our 6330 sq ft hub building contains facilities in electronics and digital manufacture, biohacking, wet-work and wood/metalwork combined with a challenge room for idea development and a large-open plan desking area. The same building also has a large cafe open to the public and garden. A short walk away from our hub building, our $>$860 sq ft Molecular Hacking space (joint with Chemistry) is now being assembled, and in our other campus is an electronics space (750 sq ft, joint with Electrical Engineering). The Advanced Hackspace hub is a single door away from an outreach oriented coummunity-makerspace. We offer and support courses, hang-outs, competitions and large-scale hackathons and provide competitive seed-corn funding. We support idea development ranging from prototyping with electronics or wood-metalwork/additive manufacture (as might be seen in an engineering-oriented makerspace \cite{Barrett, Forest}) through to hypothesis-led live-cell imaging experiments and molecular synthesis (as might be seen in an experimental laboratory in the natural sciences). `Making' is thus too narrow a term for the activities we support: idea development, with its implicit experimentation and exploration, has proved more natural. `Innovation' is also too narrow: we cover research, invention and innovation.The essence of our Advanced Hackspace is our community of members and the ideas and skills they develop and exchange Fig. \ref{figure1}.

\vspace{-3mm}

\subsection{A passport system as a tool for fast scale-up:}

\vspace{-3mm}

Our growth across spaces and in numbers can be attributed to: the enthusiasm of our founding staff; demand from university members for equipment outside their disciplines; departmental and institutional support; opening our hub space; and the early assistance of a Hackers-in-Residence programme that brought in skills gratis. A key element that allowed us to cross between departments, and so grow, was to develop a passport system. Each new space we add to our network is based on a bilateral agreement led by the departmental space-owners where, guided by precedent, they establish the kind of usage they find acceptable. Users then need to meet the safety requirements of each space in a manner akin to acquiring visas for different countries in a passport. The system is crude in that it requires repeated inductions for the different spaces and it produces heterogeneity in the kinds of access different users can have to each space. It does, however, have the virtue of being rapidly scalable without requiring a centralized mandate. Unsurprisingly, good relationships are critical for this co-use model but we found, consonant with an open-model being a stimulus, that the technicians tasked with running each university space are often our biggest supporters.

\vspace{-3mm}

\subsection{Techniques to enhance cross-disciplinary access:}

\vspace{-3mm}

It has been argued that autonomy is a primitive of health and happiness \cite{Marmot, Whitehead} but for our potential researchers to exploit the apparently free opportunities we offered we needed to remove barriers to yield equitable outcomes \cite{Philips}. We found that beyond being able to access resources for research, the other limiting factor was confidence: the self-belief required to experiment outside disciplinary comfort zones. 

University culture is oriented around progress in assessed activities, from exams to projects, and we found it vital to identify mechanisms through which our members could gain credit for undertaking autonomous projects. We exploited both Imperial's intramural courses (which offer degree-credit) and also integrated with a number of undergraduate/graduate degree courses. By these means, we could begin a culture-shift, from autonomous research being a free-time activity to being a core part of the university experience. In complement to this, and driven by demand, we shifted from operating 9-5 to operating some of our spaces 12-8pm: this allows those with courses/employment during the day to nonetheless pursue their own projects in the evening. Both integration into courses and into free-time allowed a more relaxed engagement with the opportunities we offer, thereby easing the challenge to confidence presented by our interdisciplinary framework.
 
A further challenge has been the distinction between being a facility with users and a community with members. In our early months of operation our members expected our core staff to fix their technical problems and clear jammed equipment etc.. This is rational: other university organs like libraries and gyms are run like facilities rather than communities and fee-paying college members might expect the same from the Advanced Hackspace. Since the opportunities we provide are inherently technical, our staff can never be sufficiently large to provide a facility-like level of support; further a facility expects its users to operate independently – by contrast we specifically seek to foster interdisciplinary sharing of skills and ideas in a community of researchers. Community-building in a university environment, the majority of whom will leave within 3-4 years, is a challenge. As such we have found an emphasis on co-ownership of the means of research has been particularly important: likewise community-building events like recurring hangouts, a Slack channel, our provision of a hang-out mixing-space and clear branding.  Each interaction with our members by our core team emphasises that they co-own our resource and that, unlike in a facility, they must give back as much as they take.

\vspace{-3mm}

\subsection{A catalytic role for small-scale support:}

\vspace{-3mm}

We have run twice-yearly competitions for project-support prizes of $\pounds 500-1000$. At first we preferentially offered the grants to projects linked to commercial outcomes but we discovered that some of the most promising ideas were also found in projects of a pure research character. Our criterion for success in such competitions is now  permissive: whether the idea is exciting and feasible. Our Hackers-in-Residence programme, which gave alumni free access to our spaces, in return for giving 20$\%$ of their time to support our community, had a number of extremely enthusiastic participants and was been vital in our development. We  found that, to avoid free-riding by a minority, we needed to have clear contracts.

\vspace{-3mm}

\subsection{Linking skilled interdisciplinary fellows to research:}

\vspace{-3mm}

Our fellows are highly skilled cross-disciplinary researchers that support the idea development of our community: giving technical and project advice; supporting our spaces; and delivering our courses and events. Our fellows also have the task of researching and testing/evaluating technologies and protocols which can themselves accelerate the function of our community. Examples of these range from new protocols for hackathons through ways of performing high-resolution 3d printing using low-resolution devices. Consonant with the challenge of creating a new culture we found that it was vital to have a clear onboarding process for our new fellows because the activities expected of them blending research, teaching and equipment mastery inevitably presented challenges to new starters from any one discipline.

\vspace{-3mm}

\subsection{Exploiting events for interdisciplinary mixing:}

\vspace{-3mm}

We have found that hackathons (from COPD to VoiceTech \cite{link1, link2}) have been critical for mixing across disciplines, for creating a focussed environment in which confidence in experimenting can be fostered and for bridging to industry. Others have outlined advice for the delivery of hackathons \cite{Gubin, HealthHackathon} but we found that a 3-month run-up was required for organizing a large-scale event with careful timing in the year if undergraduates are attending. Week-long showcasing events, advertising the ideas developed by our community helped members get idea feedback. From experimental seminar series we learned that: 1) that there is substantial interest in courses in mid-level technologies (from micro-fluidics to deep-sequencing) which are found in only a few undergraduate programmes and 2) that conventional lectures were poorly attended and could have been supplied by other parts of the universities. We thus moved to exclusively participatory events that required active learning and community building.

\vspace{-3mm}

\subsection{Advantages of an open culture:}

\vspace{-3mm}

As an explicitly interdisciplinary effort we have  bridged departmental barriers, however, we discovered a key trade-off which we did not anticipate in advance: between successfully exploiting the advantages of scale and becoming an exclusive monoculture. A large university must contain a diversity of efforts to realize the same outcomes and occasionally some parts of the university have replicated our efforts while other members of our university have attempted to outsource or privatise similar efforts. Our view became that we would seek to facilitate, collaborate with, and co-brand those activities that were open to the whole community, to foster and stimulate independent and analogous new efforts by others, and to disengage with efforts which actively raised barriers to access for our community.

\vspace{-3mm}

\subsection{Definition by safety  and our operating metrics:}

\vspace{-3mm}

Creating new opportunities for research for a much larger body of researchers has implications for safety: as such safety is our stated principal strategic priority. Our approach is to have safety being synonymous with the Advanced Hackspace encouraging a 
near-miss-reporting culture and drawing on best practice from elsewhere \cite{Spencer}.

We share some of our operating metrics in Table 1. We will be running bi-annual surveys of our membership and we will shortly be asking not only about members but also about their friends. This will allow us to study homophily amongst our members and to make comparisons to University members that are not actively participating in our network; in particular we will be able to identify if interdisciplinary friendships are more common among our members and to investigate gender homophily effects.

\begin{table}
  \centering
\begin{tabular}{ ||l|| } 
 \hline
 Percentage of members reporting:  \\ 
 
 that ICAH is `safe' or `very safe' \\ 
 
 \hline
 
  Percentage of members reporting:  \\ 
 
they have helped or trained other \\ 
 members. \\

\hline

 Percentage of members reporting:  \\ 

  they have formed new partnerships \\ 
  and collaborations through ICAH  \\ 
 \hline
  Gender ratio compared to university  \\

  \hline
  
Events per year \\ 
\hline

Hours of training and courses \\

\hline

Number of internal start-ups in product \\ development. \\ 

\hline

Secure applications for at least 50 \\ 
 teams towards our  grant programme \\ 

\hline

Demonstrating support of 30 minimal  \\  viable products or inventions \\ 

\hline

Fellows to develop 3 new publishable \\   protocols or technologies \\

\hline

Number of papers acknowledging ICAH  \\ 
   
   \hline

  \end{tabular}
  \caption{Operating metrics indicating our goal to create an equitable self-helping community performing research and innovation} \label{tab:sometab}
\end{table}

\vspace{-3mm}
\subsection{Discussion and the Future:}

\vspace{-3mm}

Beyond our continuing community-development and acquisition of new research technologies, we are testing new experimental programmes. The events/programmes that we currently run can cater for tens to a hundred participants; as such there is something exclusive about them: selectively investing in a few rather than the full $\sim$20,000 members of our university. We have, and will continue to experiment with, open competitions directed at all university members, including ones run with from-home components. Beyond our existing courses, we will drive the development of new `Keys to X' courses that have a different objective from introductory courses for those inside discipline X. These course will instead target those outside the discipline concerned, with the principal objective of building confidence and autonomous learning in X.

We believe there is an opportunity for academics to praxis what they preach: our courses are often tacitly designed with the idea of preparing students for a career in research and arguably universities should make this feasible for all of their students during their time at university and afterwards. By contrast, students often discover that autonomous research is something that they learn about, but are not expected to undertake either in university or in their lives afterwards. It seems sensible that universities should, instead, be beginnings: where undergraduates begin their research lives and are given confidence to pursue research throughout their lives. The idea of life-long-research, independent of being directly employed to do research, is a far more realistic proposition than it has been in the past: supported by the deskilling of technologies and the democratization of access to technology through public hackspaces and makerspaces \cite{Nesta} citizen science \cite{Silvertown} the open science and open hardware movements \cite{Landrain, Pearce} and support for science-based start-ups. There is also substantial evidence that  household innovation has a central role in our economies \cite{Sichel18, vonhippel11}.

In the past university libraries were the melting point of academic research -- cutting across all members -- we believe that resources that empower all university members to participate in a research community will constitute the university library of the future.

Acknowledgements: NJ would like to acknowledge support from the EPSRC for grants EP/K503733/1 and EP/N014529/1, NERC for the grant NE/K007270/1 and HEIF.

\end{document}